\documentstyle[aps,12pt,epsf]{revtex}
\begin{document}
\title{Interplay of pairing and multipole interactions in a simple model}
\author{Alexander Volya}
\address{
National Superconducting Cyclotron Laboratory, \\
Michigan State University,
East Lansing, Michigan 48824-1321, USA}
\date{\today}
\maketitle
\begin{abstract}
\baselineskip 14pt
The interplay of pairing and other interactions is addressed in this work
using a simple
single-$j$ model. We show that  
enhancements in pairing correlations observed through studies of  the spectra of
deformed systems,  moments of inertia, changes in transitional multipole  
amplitudes, and direct calculations of the
pairing component in the wave function, indicate that 
even without explicit matrix elements responsible for pairing, a
paired state can still 
appear from the kinematic coupling of pairing to deformation and from other
geometrical restrictions that are of extreme importance in mesoscopic systems.
Furthermore, we demonstrate that macroscopic transitions such as
oblate to prolate shape changes
can lead to strong dynamic enhancements of pairing correlations.
In this work we emphasize that the  
pairing condensate has an important dynamic and kinematic
effect on other residual interactions.
\end{abstract}
\section{Introduction}
\label{intro}
The fact that a large number of nuclei in their ground state or in their low
lying excited states are paired is supported by an overwhelming amount of
experimental evidence. This includes observation of odd-even mass difference,
appearance of a gap in the spectra,
reduction of the moment of inertia, and analog of the Josephson  effect
in pair transfer reactions. By pairing in this work we imply 
the attractive interaction between pairs of nucleons on 
time-conjugate orbitals.
It is widely accepted that the pairing interaction is responsible for
creating a superconducting paired state, however 
the realistic interaction is much more diverse
than bare pairing. The complex interplay of all interactions that  still
leads to a paired state is far from being understood.
Coherence between the pairs or even larger groups of nucleons 
can be formed in different quantum states, furthermore 
coherence may appear in the particle-hole ($p-h$) channel with
other components of interaction 
contributing to  
collective  excitations (in nuclei -
shape vibrations and giant resonances) and
deformation of the mean field. All these effects are expected to
dynamically and/or
kinematically effect the paired state.
There are also incoherent components of the interactions that 
introduce the stochastization of dynamics, but still can be influenced
by the presence of collective features in dynamics.


The appearance of a paired state is traditionally attributed to the strong
short-range residual interactions between nucleons.
However, in realistic nuclear systems all interaction matrix elements are correlated.
There is no pure pairing interaction. It is well known in the theory of superconductivity, 
and in application to nuclear structure it has been shown by Belyaev
\cite{cohfluc}  long time ago, that
an interaction with only pairing matrix elements would contradict
the fundamental principle of gauge invariance.
Recent studies
of systems with two-body random interactions \cite{mulhall,horoi01} 
indicate that survival of collective phenomena such as pairing in 
realistic systems may hinge on these correlations between different types 
of matrix elements. 
These studies have shown
that the paired ground state does  not appear in  systems governed by two-body 
random interactions;
furthermore,
even weak but uncorrelated  interactions of non-pairing type are very
destructive with respect to the pairing state.

Mesoscopic nature is yet another important property of nuclear systems. 
It has strong effect on kinematics and geometry 
of collective modes, interplay between
different excitations and phase transitions. Finiteness  was argued
to be one of the main reasons for existence of the superconducting state
in realistic nuclei \cite{kadmenskii87}. 

In this work we show that observed pairing effects in nuclei 
do  not result just from 
strong pairing matrix elements. Paired state
appears from a very complex interplay of
all residual interactions and their dynamic and kinematic behavior.
Throughout this work we use a simple single $j$-level model with only 
one species of particle in order to discuss this interplay. This model
provides  
strong kinematic constraints and the clearly pronounced
role of antisymmetry
requirements.
Pairing problem can be solved exactly in a single $j$-level and treatment
of all other interactions is  substantially simpler.  

In Sec. \ref{kin} we introduce and discuss the kinematics of interactions
in the single $j$-level model.
The main results of this work are presented in Sec. \ref{full},
where we  investigate the dynamics of paired systems, and  using a 
perturbative treatment of non-pairing interactions in the basis
of paired states \cite{EP} 
discuss  
the renormalizing effects of the pairing condensate on
other residual interactions, consider the stability of the pairing
condensate and
evaluate the applicability of the pairing-based treatment. 
We emphasize that a single-$j$ system is very kinematically constrained
and for a number of independent choices of interactions the seniority, 
the number of unpaired particles, remains conserved.
The interplay of pairing and quadrupole forces is studied in the  
``pairing plus quadrupole'' (P+Q) model. We introduce an important concept of
kinematic pairing as
specific pairing effects that appear from kinematic
restrictions present in a mesoscopic many-body system; 
they also contribute to  dynamics of nucleon-phonon interaction
\cite{kadmenskii87,kadmenskii89,barranco99}.
With numerical studies we show that those effects, ignored in the 
standard P+Q model, result in a
significant enhancement of pairing 
and directly influence the observable quantities, such as energy spectra,
moments of inertia,
and intensities of multipole transitions.
For the systems with a nearly 
half-occupied shell, where the transition from oblate to prolate deformation takes place,
pairing can be further enhanced due to the fact that in average the spherical shape tends 
to be restored in this region.

\section{Kinematics of residual interactions}
\label{kin}
The mean field is recognized as one of the most effective approaches in study of
quantum many-body systems. Along with the shape and symmetry properties of the
average many-body potential, the mean field also determines quantum numbers of
elementary excitations, the quasiparticles. 
Low-lying states in the system as
well as the response to external perturbations can be understood in terms
of the quasiparticles and their interactions, which in the lowest order
are just  pairwise collisions, see Fig. \ref{diagram}.

Spherical symmetry of the mean field is
present in many nuclei throughout the periodic table. 
With the use of a spherical basis we guarantee the exact angular momentum 
conservation, avoiding approximate projections.
Although the further discussion can be presented in a general form,
we 
limit our consideration to a single $j$-level, that is
$\Omega=2j+1$ -fold degenerate.
The general rotationally invariant two-body interaction Hamiltonian
in a single $j$-shell,
\begin{equation}
H=\sum_L V_L \sum_{\Lambda} P^{\dagger}_{L\,\Lambda} P_{L\,\Lambda}
\label{int:inth}
\end{equation}
defines the scattering of nucleon
pairs coupled to angular momentum $L\,,$
\begin{eqnarray}
\nonumber
P_{L\,\Lambda}^{\dagger}=
\frac{(-)^{L-\Lambda}}{\sqrt{2}}\,
\sum_{m_1 m_2}\, \sqrt{2L+1}\,
\left(\begin{array}{ccc}
      j& L &j\\
      m_{1}& -\Lambda &m_{2}\end{array}\right)
\,a^{\dagger}_{1} a^{\dagger}_{2}\,.
\end{eqnarray}
Rotational symmetry and Pauli antisymmetry here result in the limitation
that $L$ is even, otherwise the scattering amplitudes $V_L$ can be arbitrary. 
The $L=0$ term is responsible for pairing, the interaction of pairs on
time-conjugate orbitals,  
$|1\rangle=|j\,m\rangle$ and $|\tilde{1}\rangle=(-)^{j-m} |j\,-m\rangle\,.$
The strength of pairing is determined by $V_0\,.$

\subsection{Interactions in the particle-hole channel}

The interaction in Eq. (\ref{int:inth}) was given in the 
particle-particle ($p-p$)
channel. The
interaction can also be presented 
in the  particle-hole ($p-h$) channel.
The nucleon hole can be defined via  a canonical transformation $\hat{C}\,,$
\begin{equation}
\hat{C}\,a^\dagger_{j\,m}\,\hat{C}^{-1}  = (-)^{j-m}\,a_{j\,-m}\,\quad {\rm and}\quad
\hat{C}\,a_{j\,m}\,\hat{C}^{-1} = (-)^{j-m}\,a^\dagger_{j\,-m}\,.
\label{hole}
\end{equation}
The multipoles,
particle-hole pair states coupled to a particular angular momentum $K\,,$  are
defined as 
\begin{eqnarray}
\nonumber
{\cal M}_{K\,\kappa}=\,\sum_{m_1 m_2}\,
(-)^{j-m_1}
\left(\begin{array}{ccc}
      j& K &j\\
      -m_{1}& \kappa &m_{2}\end{array}\right)
\,a^{\dagger}_{2} a_{1}\,,
\end{eqnarray}
with the property $({\cal M}_{K\,\kappa})^\dagger=(-)^\kappa\,
{\cal M}_{K\,-\kappa}\,.$  
The lowest multipole operators with $K=0$ and $K=1$ are related to the 
constants of
motion, number of fermions $N$, and components of angular momentum operator
$J_{\kappa}\,,$
\begin{equation}
{\cal M}_{0\,0}=\frac{N}{\sqrt{\Omega}}\,,\quad {\cal
M}_{1\,\kappa}=\frac{J_{\kappa}}{\sqrt{j(j+1)\Omega}}\,,
\label{M1}
\end{equation}
where $\Omega=2j+1\,.$

The algebra of pair operators on one level is given by the following equations
\begin{equation}
[P_{L'\,\Lambda'},\, {P_{L\,\Lambda}}^\dagger]=\delta_{L\,L'}\,\delta_{\Lambda\,\Lambda'}+2\,(-)^\Lambda\,\sqrt{(2L+1)(2L'+1)}\,
\end{equation}
$$
\times \sum_{K\,\kappa}\,(2K+1)\, \left \{\begin{array}{ccc}
L&L'&K\\
j&j&j\end{array}\right\} \,\left (\begin{array}{ccc}
L&L'&K\\
-\Lambda&\Lambda'&\kappa\end{array}\right)\,{\cal M}_{K\,\kappa}^\dagger\,,
$$
\begin{equation}
[{\cal M}_{K\,\kappa},\, {P_{L\,\Lambda}}^\dagger]=-2\,(-)^\Lambda\,\sqrt{(2L+1)}\,
\sum_{L'\,\Lambda'}\,\sqrt{2L'+1}\, \left \{\begin{array}{ccc}
L'&K&L\\
j&j&j\end{array}\right\} \,\left (\begin{array}{ccc}
L'&K&L\\
\Lambda'&\kappa&-\Lambda\end{array}\right)\,P_{L'\,\Lambda'}^\dagger\,,
\end{equation}
\begin{equation}
[{\cal M}_{K\,\kappa},\, {\cal M}_{K'\,\kappa'}]=
\sum_{Q\,q}\,(1-(-)^{K+K'+Q})\,(2Q+1) (-)^q\, \left \{\begin{array}{ccc}
K&K'&Q\\
j&j&j\end{array}\right\} \,\left (\begin{array}{ccc}
K&K'&Q\\
\kappa&\kappa'&-q\end{array}\right)\,{\cal M}_{Q\,q}\,.
\label{algmult}
\end{equation}
It follows  from the last expression that the odd
multipolarity multipole operators form a closed subalgebra. Another subalgebra
relevant to pairing  is formed by operators
$P_{0\,0}\,,\,\,P_{0\,0}^\dagger\,$ and ${\cal M}_{0\,0}\,,$ it will be discussed
in detail in the following section.

\begin{figure}
\begin{center}
\epsfxsize=8.0cm \epsfbox{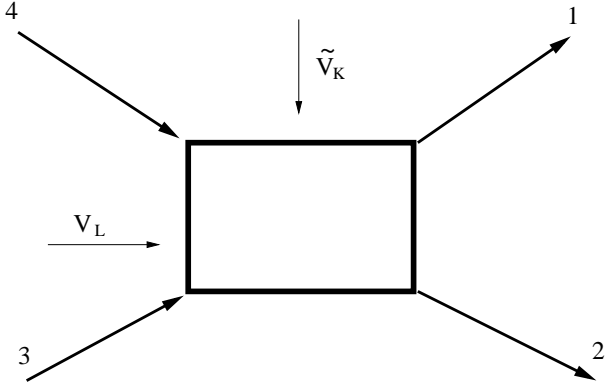}
\end{center}
\caption{Schematic diagram showing two-body scattering process.
Two  channels, particle-particle ($s$-channel) and
particle-hole ($t$-channel), are indicated by the horizontal arrow with the matrix
element $V_L$ and by the vertical arrow with the amplitude
$\tilde{V}_K\,,$ respectively, where $L$ and $K$ are total angular momenta in corresponding
channels.
\label{diagram}}
\end{figure}

The original Hamiltonian
(\ref{int:inth}) can be expressed in terms of interacting multipoles: 
\begin{equation}
H=\sum_L V_L \sum_{\Lambda} P^{\dagger}_{L\,\Lambda} P_{L\,\Lambda}= 
\epsilon N -\sum_K \frac{\tilde{V}_K}{2} \sum_{\kappa} {\cal M}^{\dagger}_{K\,\kappa} 
{\cal M}_{K\,\kappa}\,.
\label{FULL:degenerate}
\end{equation}
The transformation formulas between the 
$p-p$ and $p-h$ channels on one
level are
\begin{equation}
\tilde{V}_K=(2K+1)\sum_L\,(2L+1)\,
\left\{\begin{array}{ccc}
j&j&L\\
j&j&K\end{array}\right\} V_L\,,
\label{onel_vk}
\end{equation}
\begin{equation}
V_L=\sum_K\,
\left\{\begin{array}{ccc}
j&j&K\\
j&j&L\end{array}\right\} \tilde{V}_K\,,
\label{onel_vl}
\end{equation}
\begin{equation}
\epsilon=\frac{1}{2\Omega}\,\sum_K\,\tilde{V}_K\,.
\label{spenergy}
\end{equation}
This transformation, often attributed to Pandya \cite{pandya56},
was first seriously discussed from the viewpoint of underlying physics
and practically used by Belyaev 
\cite{belyaev60}, see also  \cite{de-shalit63,bertsch72}.
Schematically the transformation from the $p-p$ to $p-h$ channel is shown in
Fig. \ref{diagram}.

Fermionic antisymmetry
requires  that pairs of fermions on one level  couple to even
angular momentum, therefore 
interaction (\ref{int:inth}) is
defined by $j+1/2$ independent parameters $V_L$ with $L=0,2,\dots 2j-1\,.$
This fact is obscured in the Hamiltonian in the $p-h$ channel, where 
particle and hole can couple to any angular momentum. The number of independent
parameters is still the same, however instead of a simple limitations $V_L=0$ 
for $L=1,3,\dots 2j$ in the $p-h$ channel the constraints for $\tilde{V}_K$
are given by linear conditions
\begin{equation}
\tilde{V}_K=(2K+1)\sum_{K'}\,(-)^{K+K'}\,
\left\{\begin{array}{ccc}
j&j&K'\\
j&j&K\end{array}\right\} \tilde{V}_{K'}\,.
\label{onel_rel}
\end{equation}
Analogous relations are known in the macroscopic Fermi-liquid theory.

It is convenient to 
introduce a projection operator $\hat{\Theta}$ that acting on
$V_L$ projects out only its physical component,
\begin{equation}
\hat{\Theta}\,V_L=\frac{1+(-)^L}{2}\,V_L\,.
\label{projPP}
\end{equation}
A similar operator also
exists in the space of $\tilde{V}_K\,,$ which is
just a linearly transformed set of interaction parameters, see Eq.
(\ref{onel_vk}). However, it is no longer diagonal:
\begin{equation}
\hat{\Theta}\,\tilde{V}_K=\frac{1}{2}\,\tilde{V}_K+ \frac{1}{2}\,(2K+1)\sum_{K'}\,(-)^{K+K'}\,
\left\{\begin{array}{ccc}
j&j&K'\\
j&j&K\end{array}\right\} \tilde{V}_{K'}\,.
\label{projQQ}
\end{equation}
The condition that all $V_L$ with odd $L$ vanish is equivalent to
$\hat{\Theta} V_L=V_L\,,$ similarly in the $p-h$ channel
Eq. (\ref{projQQ}) is equivalent to $\hat{\Theta} \tilde{V}_K=\tilde{V}_K\,.$

Eq. (\ref{onel_rel}) can be viewed as an eigenvalue
equation, where the kernel can be brought to a symmetric form
by a simple rescaling of ${\tilde V}_K$ by $(2K+1)^{1/2}$.
This eigenvalue problem can
be resolved by separating the 6-$j$ symbol in the kernel
with the recoupling technique. This, though, does not lead to anything new,
because as a result one obtains that among $2j+1$ eigenvalues there are
$j+1/2$ eigenvalues that are zero and the same number of eigenvalues
equal to one, which is a consequence of this operation being a projection. 
Any physical eigenmode for the set of
${\tilde V}$ corresponds to one particular $V_L$, and can be constructed
using Eq. (\ref{onel_vk}), since the projection operator is diagonal in the 
$p-p$ channel. 

The special cases of Eqs. (\ref{onel_rel}) and (\ref{onel_vk}) 
result in 
\begin{equation}
\sum_K\,\tilde{V}_K=\sum_L(2L+1) V_L = -\Omega \tilde{V}_0\,,\quad
\sum_K\,(-)^K\,\tilde{V}_K=-\Omega V_0\,,
\label{A:zeroM}
\end{equation}
and effective single-particle energy in (\ref{spenergy}) can be written as
\begin{equation}
\epsilon=\frac{1}{2\Omega}\,\sum_L(2L+1) V_L=-\frac{\tilde{V}_0}{2}
=\frac{1}{2\Omega}\,\sum_K\,\tilde{V}_K\,\,.
\label{spe}
\end{equation}
These constraints are usually not addressed in nuclear models, 
because the Hamiltonian given in the $p-h$ form is still good even if they 
are not satisfied, it merely contains 
the components that
identically vanish in any fermionic many-body state, and still only $j+1/2$
independent combinations of parameters determine the interaction. 
The arbitrary amount of these components make the $p-h$ form of the 
Hamiltonian expressing
the same interaction not unique. The unique form can be reached if 
all non-physical components are removed with
projection operators (\ref{projPP}) or (\ref{projQQ}). After projection
the interaction remains physically identical to the original one, but the
new parameters satisfy
Eq. (\ref{onel_rel}). 

The situation can be illustrated by an example 
of monopole interaction, where all nucleon pairs interact with identical 
strength, $V_L=1$ for all even $L\,,$
\begin{equation}
H=\sum_{L=0,2,\dots 2j-1} \sum_{\Lambda} P^{\dagger}_{L\,\Lambda} P_{L\,\Lambda}\,.
\label{mod00}
\end{equation}
This interaction is very simple because its effect is only in 
counting the number
of particle pairs in the system, therefore all states have
the same energy 
\begin{equation}
E=\frac{N(N-1)}{2}\,.
\label{monomo}
\end{equation}

Going to the  $p-h$ channel Eq. (\ref{onel_vk}), we rewrite (\ref{mod00}) 
in the form of interacting multipoles
\begin{equation}
H=\epsilon N -\sum_K \frac{\tilde{V}_K}{2} \sum_{\kappa} {\cal M}^{\dagger}_{K\,\kappa} 
{\cal M}_{K\,\kappa}\,
\end{equation} 
where 
\begin{equation}
\tilde{V}_K=-\frac{\Omega}{2}\,\left ( \delta_{K\, 0} -
\frac{2K+1}{\Omega}
\right )\,,\quad {\rm and } \quad \epsilon=\frac{\Omega-1}{4}.
\label{min}
\end{equation}
All components of this interaction respect Pauli principle and 
Eq. (\ref{onel_rel}) is  fulfilled.
However, it is not obvious
that the action of this Hamiltonian is equivalent to counting the 
particle pairs.
In order to gain a simpler form we add to this Hamiltonian a non-acting part
\begin{equation}
H'= H+\sum_{L=1,3,\dots 2j} \sum_{\Lambda} P^{\dagger}_{L\,\Lambda} P_{L\,\Lambda}\,.
\end{equation}
Transforming $H'\,,$ where all
$V_L=1\,,\quad L=0,1,2,\dots 2j\,,$
to the $p-h$ channel,  we get
\begin{equation}
\tilde{V}_{K}=-\Omega\,\delta_{K\,0}\,,\quad {\rm and} \quad \epsilon=-\frac{1}{2}\,.
\end{equation}
Thus, only the monopole term is present in this interaction and 
\begin{equation}
H'=\frac{N(N-1)}{2}\,.
\label{NN}
\end{equation}
Although Hamiltonians $H$ and $H'$ 
have a very different form in the $p-h$ channel, they are identical
in their action on a physical state. Despite the fact that 
introduction of inactive components may allow for a 
simpler form of the Hamiltonian, 
the form where Eq. (\ref{onel_rel}) is satisfied,  is preferred, not
only because it allows to define interaction in the unique way, 
but also because it explicitly 
shows couplings between different physical $p-h$ excitations by virtue of
Eq. (\ref{onel_rel}).  

The role that each interaction parameter $V_L$ or $\tilde{V}_K$
plays in determining the state
of a many-body system is very complex, and generally for realistic systems
these parameters
are correlated by their common physical origin (such as core polarization or
meson exchange for example)
beyond the previously 
discussed kinematic restrictions. 
In realistic systems
there are some $V_L$, $\tilde{V}_K\,,$ and possibly their certain
linear
combinations, 
that have a significant tendency
to form  nuclear states with special coherent
properties and symmetries. 
The 
$V_0\,,$ pairing matrix element $V_0$, is known to be responsible for
collective and macroscopic coherent effects similar to superconductivity and
superfluidity in large many-body fermionic systems. 
Similarly, in the particle-hole channel $\tilde{V}_2$ plays an important
role for formation of collective vibrations and quadrupole deformation.

\subsection{Particle-hole symmetry}
Residual interactions can be formulated in the hole-hole (h-h) channel.
With the transformation in Eq. (\ref{hole}) pair operators transform as
\begin{equation}
\hat{C}\,P^\dagger_{L\,\Lambda}\,\hat{C}^{-1} = P_{L\,-\Lambda}\,(-)^{1-\Lambda}\,,\quad
{\rm and}\quad \hat{C}\,{\cal M}^\dagger_{K\,\kappa}\,\hat{C}^{-1}=(-)^{1+K}\,{\cal
M}^\dagger_{K\,\kappa}\,,\quad K\ne 0\,.
\end{equation}
The number operator transforms as
\begin{equation}
\hat{C}\,N\,\hat{C}^{-1} = \Omega-N\,.
\end{equation}
With the help of the 
identity
\begin{equation}
\sum_{\Lambda}\,[P_{L\,\Lambda},\, {P_{L\,\Lambda}}^\dagger]=
(2L+1)\left (1-\frac{2N}{\Omega} \right )\,, 
\end{equation}
the Hamiltonian in Eq. (\ref{int:inth}) 
can be transformed to the hole-hole representation
\begin{equation}
\hat{C}\,H\,\hat{C}^{-1}=\sum_L V_L \sum_{\Lambda} P^{\dagger}_{L\,\Lambda}
P_{L\,\Lambda}-\tilde{V_0}(\Omega-2N)\,.
\label{int:hamhh}
\end{equation}
Since the
number of particles (or holes) is a constant of motion, the $p-p$ to $h-h$
transformation simply results in a constant shift of energy,
while leaving the interaction invariant. The same result can be traced using
the multipole-multipole ($p-h$) representation of interactions. Here
all multipole terms with $K\ne 0$ are invariant, and any changes are due to the 
monopole and single-particle terms.
Particle-hole invariance results in important consequences \cite{lawson}: 
an expectation value of 
any odd multipole moment of the $N$-particle system is equal to the multipole
moment of the corresponding state in the  $\Omega-N$ system;
any even multipole
moment is equal in magnitude but  has an opposite sign in the corresponding states
of the $p-h$ conjugate system.
In particular, the particle-hole symmetry requires that 
expectation values of all even multipole moments 
identically vanish in the half-occupied 
shell. Therefore, a half-occupied shell can not be deformed. 
As we further show, this
effect turns out to be  helpful for preserving a paired
state. This kinematic suppression of
deformation is a result of a
phase transition on the mean-field level from oblate to prolate deformation. 
Similar to the pairing phase
transition, the 
mesoscopic nature of the system smoothens the sharp changes, thus extending 
the region of large fluctuations and
suppressed deformations. 

\section{Pairing and other interactions}
\label{full}
\subsection{Pairing interactions and degenerate model}
\label{quasispin}
The first steps towards understanding the nucleon pairing
were taken even before Bardeen, Cooper and Schrieffer developed
their powerful BCS method \cite{BCS57} in 1957.
The degenerate model involves a single degenerate
single-particle level.
The algebraic properties involving $P\equiv P_{0\,0}\,,$ 
$P^{\dagger}\equiv P^\dagger_{0\,0}\,,$ and ${\cal M}\equiv{\cal M}_{0\,0}$ operators 
on one level are particularly simple: 
\begin{equation}
\left [ {\cal M}_{K \kappa},\, P^{\dagger}\right ]=\frac{2 (-)^\kappa 
}
{\sqrt{\Omega(2K+1)}}\,P^{\dagger}_{K\,-\kappa}\,
\label{CommEVK}
\end{equation}
for even $K\,,$ and
\begin{equation}
\left [ {\cal M}_{K \kappa},\, P^{\dagger}\right ]=0\,
\label{CommODK}
\end{equation}
for odd $K\,,$
\begin{equation}
\left [ P_{L \Lambda},\, P^{\dagger}\right ]=\delta_{L\,0}\,\delta_{\Lambda\,0}  - 
\frac{2 \sqrt{2L+1}}
{\sqrt{\Omega}}\,{\cal M}_{L \Lambda}\,.
\end{equation}
The important $L=0$ case, 
\begin{equation}
\left [ P,\, P^{\dagger}\right ]=1 - 
\frac{2}
{\sqrt{\Omega}}\,{\cal M}=1-\frac{2N}{\Omega}\,,
\end{equation}
shows that the zero spin set of operators ($P\,,$ $P^\dagger\,,$ and ${\cal M}$) 
form the SU(2) algebra. 
By defining a quasispin ${\vec {\cal L}}\,,$
\begin{equation}
{\cal L}_z=\frac{N}{2}-\frac{\Omega}{4}\,,
\quad {\cal L}^{+}=\sqrt{\frac{\Omega}{2}}\,P^{\dagger}
\,,\quad {\cal L}^{-}=\sqrt{\frac{\Omega}{2}}\,P\,,
\label{qspin}
\end{equation}
we can satisfy the  commutation relations. The pure pairing interaction
preserves quasispin; this can be converted into conservation of
seniority, the number of unpaired 
particles $s$.
This is the cornerstone of Kerman's \cite{kerman61} method 
and the EP algorithm \cite{EP} of exact solution of the pairing problem for the realistic
level scheme. 
The eigenvalues  of ${\cal L}_z$ and ${\vec {\cal L}}^2={\cal L}({\cal L}+1)$
are related to the particle number $N$ and seniority $s$ according to
\begin{equation}
{\cal L}_z=\frac{N}{2}-\frac{\Omega}{4}\,,\quad {\cal L}=\frac{\Omega}{4}-
\frac{s}{2}\,.
\end{equation}

By repeatedly commuting pair operators we get  
\begin{equation}
\left [P,\,\left (P^{\dagger} \right )^{n}\right ]=
\left (P^{\dagger} \right )^{n-1}\, n \left 
(1-\frac{2n-2}{\Omega}-\frac{2N}{\Omega}\,
\right )\,,
\end{equation}
\begin{equation}
\left [ P_{L\,\Lambda},\, \left (P^\dagger \right )^n \right]=
-\frac{2n(n-1)}{\Omega} \, \left(P^{\dagger}\right)^{n-2}\, (-)^\Lambda P^{\dagger}_{L\,-\Lambda}
-\frac{2n\sqrt{2L+1}}{\sqrt{\Omega}}\left(P^{\dagger}\right)^{n-1}
{\cal M}_{L \Lambda}\,,\,\,L\ne 0\,,
\label{comm_PP}
\end{equation}
and
\begin{equation}
\left [M_{K\, \kappa},\, \left (P^{\dagger} \right )^n \right ]=
\left \{ \begin{array}{l}
\frac{2 n 
}
{\sqrt{\Omega(2K+1)}}\,\left(P^{\dagger}\right )^{n-1}\,
(-)^\kappa P^{\dagger}_{K\,-\kappa}\,,\quad K\,\,{\rm even}\\
0\,,\quad K\,\,{\rm odd}
\end{array}\right .\,.
\label{comm_MP}
\end{equation}    
The last expression results from $[[M_{K\, \kappa},\, P^{\dagger} ],\, P^{\dagger}]=0\,.$

Collective paired states (the condensate) can be built
on any  state $|N=s,\,s\rangle$ with $s$ unpaired particles by the simple action of
the pair creation operators $\left (P^{\dagger}\right )^{n}|s,\,s\rangle$, 
resulting in the state with  $n$  pairs in a condensate 
$|N=2n+s,\,s\rangle \,.$
The normalization of such a state can be obtained using the momentum algebra,
or iteratively with the help of the commutation relations,  
\begin{equation}
\langle N=s,\,s|P^n \left (P^{\dagger} \right )^{n}|N=s,\,s \rangle =
 \frac{[(\Omega/2)-s] !\, n! }{(\Omega/2)^n\, [(\Omega/2) -n-s]!}\,
\langle N=s,\,s|N=s,\,s \rangle\,.
\label{norms0}
\end{equation}
The seniority formalism is useful because it
takes all unpaired states $|s,s\rangle $  as a foundation upon which all
other states are uniquely built by adding a paired condensate. 
The simplest lowest non-zero seniority states are 
the $s=1$ state  $|N=1,\,1 (j m)\rangle = a^{\dagger}_{j m}|0\rangle$
and 
the $s=2$ state $(P_{L\ne 0\,\Lambda})^{\dagger}|0\rangle=|2,\,s=2 (L \Lambda)\rangle$,
both of which are normalized to unity with our definitions.

\subsection{Pairing-based treatment on non-pairing residual interactions}

In this subsection we will assume that the system is paired,  
the ground state has seniority $s=0$ (assuming even $N$), 
and the lowest excited state has $s=2\,.$ Using 
the
paired states we
will evaluate the contribution of all residual interactions to the
energy, the EP+monopole method \cite{EP}; 
using the states with $s=2$ we will discuss the behavior of two
unpaired nucleons
in the presence of the $N-2$-particle condensate, and address the validity of 
the initial assumption that in the presence of all residual interactions 
the system is still paired.

In the lowest order of perturbation theory for the $s=0$ state  we have to
examine the expectation values of all terms in the Hamiltonian of
Eq. (\ref{FULL:degenerate}) for the paired state $|N,s=0\rangle\,.$  
Following the commutation relations (\ref{comm_MP}) we obtain
\begin{equation}
\sum_{\kappa}\langle N,\, 0|{\cal M}^{\dagger}_{K\,\kappa} 
{\cal M}_{K\,\kappa} |N,\, 0 \rangle=\frac{2N(\Omega-N)}
{\Omega(\Omega-2)}\,,\quad K=2,4,\dots\,,
\end{equation}
where only even multipoles contribute.
The $K=0$ case is proportional to the square of 
the particle number
 \begin{equation}
\langle N,\, 0|{\cal M}^{\dagger} 
{\cal M} |N,\, 0 \rangle=\frac{ N^2}{\Omega}\,.
\end{equation}
Using  (\ref{comm_PP})
it can be shown that
\begin{equation} 
\langle N,\, 0|P^{\dagger}_{L\,\Lambda} 
P_{L\,\Lambda} |N,\, 0 \rangle = \frac{N(N-2)}{\Omega (\Omega-2)}\,,\quad
L\ne 0\,.
\label{PP_one}
\end{equation}
The $L=0$  corresponds to the solution of pairing in
the degenerate model,
\begin{equation}
\langle N,\, s|P^{\dagger} 
P |N,\, s \rangle = \frac{N-s}{2\Omega} \, (\Omega-N-s+2)\,,
\label{degeneratesolution}
\end{equation}
where the number of particles in the state is $N=2n+s\,.$ This expression is
valid for any seniority $s\,.$
The expectation value of the Hamiltonian in the paired state is thus
$$
\langle N,\, 0|H|N,\, 0 \rangle =
\frac{V_0}{2}\, \frac{N(\Omega-N)}{\Omega-2} + \frac{N(N-2)}{\Omega (\Omega-2)} \sum_L
(2L+1) V_L
$$
\begin{equation}
=\frac{V_0}{2} \frac{N(\Omega-N)}{\Omega-2}-\tilde{V}_0\,\frac{N(N-2)}{\Omega-2}\,.
\label{FULL:s0}
\end{equation}
The same result can be obtained in the multipole channel.
The result (\ref{FULL:s0}) is of particular interest since here the exact
expectation value of the full Hamiltonian on the paired wave function is
just the sum of the pairing and monopole contributions. The treatment of energy
within the ``exact pairing plus monopole'' (EP+M$_{0}$) approximation 
is therefore the lowest order
perturbation treatment \cite{EP}.
It is also important to mention that $s=4\,$ is the lowest
seniority mixed with $s=0$ by the 
non-pairing part of the Hamiltonian.
This is related to the conservation 
of angular momentum. The state of two unpaired particles cannot
have spin zero since two particles on a single $j$-level 
have only one spin zero state which belongs
to seniority zero. 

To consider the states of higher seniority it is convenient to utilize the
quasispin group properties.
All states with a given
seniority have the same  expectation value of ${\bf {\cal L}}^2$ and the
number of particles in the paired condensate is 
reflected only in the  quasispin projection
${\cal L}_z\,.$ 
With the help of the SU(2) quasispin group all operators can be classified by
their seniority selection rules, and any expectation value
$\langle N,s |X|N', s'\rangle$ can be related to a quasispin-reduced
matrix element 
$\langle s ||X||s'\rangle\,$ using the Wigner-Eckart theorem. 
These procedures are discussed in
detail by Talmi \cite{talmi}. 
From the previously discussed
commutation relations it follows that for odd $L\,,$ ${\cal M}_{L\,\Lambda}$ 
is a quasispin scalar, while the even-$L$  pair operators 
$P^{\dagger}_{L\,\Lambda}\,,$
$P_{L\,\Lambda}\,,$ and ${\cal M}_{L\,\Lambda}$ can be combined in  components
of quasispin vectors,
for $L=0$ they  define quasispin via Eq. (\ref{qspin})
The Hamiltonian (\ref{FULL:degenerate})
is a mixture of scalar, vector and second
rank tensor in quasispin space,
$$
H={\cal H}_0+{\cal H}_1+{\cal H}_2\,.
$$
With the aid of  the multipole expansion, the
components of the Hamiltonian can be
explicitly extracted. 
Due to the symmetry properties, the product of two identical 
vectors cannot have a vector component, because the cross product of
two equal vectors is identically zero.
Therefore ${\cal M}^{\dagger}_{K\,\kappa}
{\cal M}_{K\,\kappa}$ with non-zero even  values of $K\,$ contain no
quasivector component. 
Thus,   
the quasivector part is fully contained in the $K=0$ terms
\begin{equation}
{\cal H}_1=-\tilde{V}_0 \left ( N -\frac{\Omega}{2}\right )=- 2 \tilde{V}_0 {\cal L}_z\,.
\label{H1}
\end{equation}
The quasispin-quadrupole parts, that are only present in terms 
$
{\cal M}^\dagger_{K\,\kappa}{\cal M}_{K\,\kappa}$
with even $K\,,$ can be separated by decomposing the product, for example
\begin{equation}
({\cal L}^z)^2=\underbrace{\frac{1}{6}\left (2({\cal L}^z)^2+
{\cal L}^{+}\,{\cal L}^{-}+{\cal L}^{-}\,{\cal L}^{+}\right )}_{\rm scalar}+
\underbrace{\frac{1}{6}\left (4({\cal L}^z)^2-
{\cal L}^{+}\,{\cal L}^{-}-{\cal L}^{-}\,{\cal L}^{+}\right )}_{\rm quadrupole}\,.
\end{equation}
Therefore
\begin{equation}
{\cal H}_2=\frac{1}{3}\,\sum_L \left ( V_L + \frac{2 \tilde{V}_L}{2L+1} 
\right ) \sum_{\Lambda} P^{\dagger}_{L\,\Lambda}
P_{L\,\Lambda}\,-\,\frac{V_0+2\tilde{V}_0}{3} \left (\frac{\Omega}{4}- N\right ) 
\label{H2}\,,
\end{equation}
and the remaining part is a quasiscalar
\begin{equation}
{\cal H}_0=\frac{2}{3}\,\sum_L \left ( V_L - \frac{\tilde{V}_L}{2L+1} 
\right ) \sum_{\Lambda} P^{\dagger}_{L\,\Lambda}
P_{L\,\Lambda}\,+\,\frac{V_0}{3} \left (\frac{\Omega}{4}-N \right ) 
+\,\frac{\tilde{V}_0}{3} \left (N-{\Omega} \right ) 
\label{H0}\,.
\end{equation}
The quasivector part is proportional to ${\cal L}_z$ and can act only
within a multiplet, generating no change in seniority.
Therefore in all transitions generated by the Hamiltonian and leading to a
change in seniority the quadrupole part ${\cal H}_2$ is the only active
component, changing seniority by either 2 or 4 units. 
Using the Wigner-Eckart theorem for transitions generated by the
second rank tensor in seniority, we obtain
\begin{equation}
\frac{\langle N,s|H|N,s-4 \rangle}{\langle s,s|H|s,s-4 \rangle} = \frac{1}{2}\sqrt{\frac{(N-s+4)(N-s+2)(\Omega-N-s+2)(\Omega-N-s+4)}{2(\Omega-2s+2)(\Omega-2s+4)}}\,,
\end{equation}
\begin{equation}
\frac{\langle N,s|H|N,s-2 \rangle}{\langle s,s|H|s,s-2 \rangle} = \,
\frac{\Omega-2N}{\Omega-2s}\,\sqrt{\frac{(N-s+2)(\Omega-N-s+2)}{2(\Omega-2s+2)}}\,.
\end{equation}
The situation with the diagonal in seniority contribution is somewhat more
difficult since
both components, quasiscalar and second rank tensor in quasispin space, 
are active in this case
\begin{equation}
\langle N,s (\xi) |H|N,s (\xi')\rangle=\langle s,s (\xi)|H|s,s
(\xi')\rangle-\tilde{V}_0(N-s)\delta_{\xi \xi'}-
\label{EPN}
\end{equation}
$$
\frac{6(N-s)(\Omega-N-s)}{(\Omega-2s)(\Omega-2s-2)}\,
\langle s,s (\xi)|{\cal H}_2|s,s
(\xi')\rangle\,;
$$
$\xi$ here denotes all other quantum numbers not related to quasispin
which are needed to identify the state.
It can be  convenient to extract a quadrupole component using two
states in the quasispin multiplet:  a state with
no paired particles and a state with  the same seniority but with
one condensate  pair.
It can than be shown \cite{talmi} that 
\begin{equation}
\langle N,s (\xi) |{\cal H}_2|N,s (\xi')\rangle=
\frac{\Omega-2s}{12}\,
\left (\langle s+2,s (\xi) |H|s+2,s (\xi') \rangle-\langle s,s (\xi) |H|s,s
(\xi')\rangle
+2\tilde{V}_0 \delta_{\xi \xi'} \right )\,.
\end{equation}

The previously obtained formula for the $s=0$ case, Eq. (\ref{FULL:s0}),
results from the following
conditions
\begin{equation}
\langle N=0,s=0|H|N=0,s=0\rangle =0\,,\quad \langle N=0,s=0|{\cal H}_2|N=0,s=0\rangle=
-\frac{\Omega(V_0+2 \tilde{V}_0)}{12}\,.
\end{equation}
The $s=1$ expression follows directly from Eq. (\ref{EPN}) supplemented with
\begin{equation}
\langle 1,1 (j m)|H|1,1 (j m)\rangle =0\,,\quad
\langle 1,1 (j m)|{\cal H}_2|1,1 (j m)\rangle =-\frac{(\Omega-4)(V_0+2 \tilde{V}_0)}{12}\,,
\end{equation}
\begin{equation}
\langle N,1 (j m)|H|N,1 (j m)\rangle=\frac{N-1}{\Omega-2}\left (
(\Omega-N-1) \frac{V_0}{2} - 
(N-1)
\tilde{V}_0
\right )\,.
\label{FULL:s1}
\end{equation}
The answer here contains only the pairing and monopole terms. 
An extra particle influences the pairing condensate only through the Pauli blocking.

Two unpaired particles above the pairing condensate behave  differently,
and their interaction is strongly renormalized. For the $s=2$ case we get
\begin{equation}
\langle N,2 (J M)|H|N,2 (J M)\rangle=V_J-(N-2) \tilde{V_0}
+
\label{FULL:s2}
\end{equation}
$$
\frac{(N-2)(\Omega-N-2)}{(\Omega-6)(\Omega-4)}\left\{ (\Omega-8)
(\frac{V_0}{2}+\tilde{V}_0)-
2V_J-\frac{4\tilde{V}_J}{2J+1}\right \}\,.
$$
This equation shows that unpaired particles interact in the channel with
angular  momentum
$L$ with a reduced strength
\begin{equation}
V^{\prime}_{L}= V_{L}\,\left (1-2\frac{N_{\rm p}(\Omega-N_{\rm
p}-4)}{(\Omega-6)(\Omega-4)}\right )\,\,,\quad N_{\rm p}=N-s\,,
\label{FULL:renorm}
\end{equation}
because of the presence of the $N_{\rm p}$-particle condensate. The reduction is
proportional to the expectation value of $P^\dagger P$ and
has a parabolic dependence on $N_{\rm p}$,
resulting in the maximum weakening
of the pair interaction by about a factor of
$1/2\,.$
Addition of two unpaired particles implies an extra blocking of pairing
and therefore requires more energy as compared to the case of the
two particles being paired. The additional energy  comes from a
two-quasiparticle excitation and 
is proportional to $V_0\,.$  The nontrivial contribution
from other interactions also enters  through the monopole and $\tilde{V}_J$
terms.

To investigate the stability of the paired state we
consider the separation energy of a particle pair from the condensate
$S=E(N,s=0)-E(N,s=2)\,.$  In the approximation of large $\Omega\,$ and $N_{\rm
p} \,,$
Eqs. (\ref{FULL:s0}) and (\ref{FULL:s2}) give
\begin{equation}
S=V_0-V_L+2\left ( V_L+\frac{2\,\tilde{V}_L}{2L+1} \right ) \,\frac{N_{\rm p}}{\Omega}\,\left (1-\frac{N_{\rm p}}{\Omega}
\right )=V_0-V^{\prime}_L+4\,\frac{\tilde{V}_L}{2L+1}\,\frac{N_{\rm p}}{\Omega}\,\left (1-\frac{N_{\rm p}}{\Omega}
\right )\,.
\label{FULL:separation}
\end{equation}
The self-consistency of this treatment based on pairing requires that
pairing be stable and $S<0\,.$  
Eq. (\ref{FULL:separation}) as a function of the condensate size $N_{\rm
p}/\Omega $ has three extremum points. The two points at the edges of occupancy
$N_{\rm p}/\Omega=0$ and $N_{\rm p}/\Omega=1$ are equivalent due to
particle-hole symmetry and result in the obvious condition
\begin{equation}
V_0<V_L\,,\,{\rm for\,\, any}\,\,L\ne 0\,. 
\label{cond1}
\end{equation}
A non-trivial condition appears in the third point of extremum, for the half-filled
shell $N_{\rm p}/\Omega=1/2\,,$ where
\begin{equation}
V_0<\frac{V_L}{2}-\frac{\tilde{V_L}}{2L+1} \,\,{\rm
for\,\, any\,\,even}\,\,L\ne 0\,.
\label{cond2}
\end{equation} 

Quenching of residual matrix elements in the $p-p$ channel, according to 
Eq. (\ref{FULL:renorm}) is an important
phenomenon, which can prevent non-pairing interactions, especially ones that are incoherent
with respect to the mean field deformation, from
destructing the paired state. 
However, as can be seen from Eq. (\ref{cond2}),
the multipole-multipole correlations can damage the
pairing state, and the above pairing-based treatment
may become inappropriate. In  Sec. \ref{SECPQ} we will continue 
the discussion of interplay of pairing and coherent multipole modes.      

\subsection{Seniority conservation and kinematics of interactions}
The one level model is very restrictive kinematically, and constraints are
somewhat favoring pairing. Although the interaction on a single level 
is defined with the explicit use of $j+1/2$ 
independent parameters, such as  $V_L$ with even $L\,,$   
there are only few independent linear
combinations that result in a seniority mixing interaction. Besides an obvious
pairing component $V_0\,,$ it is also possible to 
find a number of interactions that produce no quadrupole part ${\cal H}_2=0\,.$
It follows from Eq. (\ref{H2}) that this will happen if 
for an arbitrary even $L$
\begin{equation}
V_L+\frac{2\tilde{V}_L}{2L+1}=0\,.
\label{nomix}
\end{equation}
This condition results in linear equation
\begin{equation}
\sum_{L'}\,{\cal K}_{L\,L'}\,V_{L'}=0\,,\quad {\rm where}\quad 
{\cal K}_{L\,L'}=\delta_{L\,L'}+(2L'+1) \,
\left\{\begin{array}{ccc}
j&j&L'\\
j&j&L\end{array}\right\}\,.
\end{equation}
Since 
\begin{equation}
{\cal K}^2=3 {\cal K}\,,
\end{equation}
the kernel ${\cal K}$ has only two different eigenvalues, 3 and 0. All eigenvectors 
corresponding to the zero eigenvalue of ${\cal K}$ are independent solutions of
Eq. (\ref{nomix}).
Solutions of Eq. (\ref{nomix}) 
can be obtained as
\begin{equation}
V_L=\left ( \begin{array}{ccc}
j&j&L\\
{ m}&-{ m}&0\end{array}\right )^2\,,\quad L=0,2,\dots 2j-1\,.
\label{generate}
\end{equation} 
The corresponding parameters in the $p-h$ channel are 
\begin{equation}
\tilde{V}_K=\frac{2K+1}{2} \left \{(-)^K\, \left ( \begin{array}{ccc}
j&j&K\\
{ m}&-{ m}&0\end{array}\right )^2- \left ( \begin{array}{ccc}
j&j&K\\
{ m}& { m}& -2{ m}\end{array}\right )^2 \right \}\,,\quad K=0,1,\dots
2j\,.
\label{generate2}
\end{equation}
In the above equation the second term in the brackets identically vanishes for
all even values of $K\,.$
It is clear that Eq. (\ref{nomix}) is
satisfied by (\ref{generate}) and (\ref{generate2}).
Not all solutions generated by $j+1/2$ different values of $m=1/2,\,3/2,\,\dots j$ are linearly 
independent, this in general allows for existence of some independent linear 
combinations of interaction parameters $V_L$ that result in seniority mixing Hamiltonians.
The number of linearly independent solutions of Eq. (\ref{nomix}) is 
$2 k+1-\delta_{r\,0}\,,$ where $k$ and $r$ are 
determined 
as $j+1/2=3k+r\,,$ where r=0,1,2 is the residue. (\ref{generate}) 
generates all these solutions, they correspond to the zero eigenvalue
of the kernel ${\cal K}\,,$ and result in Hamiltonians that preserve seniority. 
Furthermore, since the quasiscalar ${\cal H}_0$ and quasivector ${\cal H}_1$ parts of
the interaction result in a trivial $N$-dependence of the spectra as follows from 
(\ref{EPN}),
\begin{equation}
E(N,s (\xi) )=E(N=s,s (\xi))-\tilde{V}_0(N-s)\,,
\end{equation}  
the relative spacings
between states of the same quantum numbers including seniority are independent of $N$ for 
interactions that satisfy Eq. (\ref{nomix}).

As a remark we note that the $\delta$-interaction can be generated by
Eq. (\ref{generate}) with ${ m}=1/2$ and thus conserves seniority \cite{lawson}.
In addition to all these $(2 k+1-\delta_{r\,0})$ choices, there is 
one trivial and linearly independent (for $j>1/2$) freedom of 
selecting $V_0$ that also results in a seniority conserving Hamiltonian.
To emphasize these kinematic limitations  
we just mention that for $j\le 7/2$ any interaction conserves seniority and 
as it has been recently
noticed in \cite{rowe01} for $j=19/2\,,$ for example, 
out of 9 possible independent choices of parameters
only 2 lead to seniority non-conservation. In the limit of large $j$ only about
one third of parameters result in seniority mixing Hamiltonians.
Finally, for the linearly independents sets of interaction parameters corresponding to 
the eigenvalue 3, the
resulting Hamiltonian is almost purely quadrupole in quasispin
\begin{equation}
{\cal H}_2=H-V_0\left ( \frac{\Omega}{4}-N \right )\,,\quad {\cal H}_0=-\frac{V_0\,\Omega}{4}\,.
\end{equation} 
The fact that it is possible to find non-zero sets of interaction parameters
that result in  a vanishing  quadrupole in quasispin component of the Hamiltonian 
is non-trivial, it is furthermore interesting that the number of independent 
possibilities is large. This is a result of a very restrictive kinematics of
interactions on a single $j$-level.  

For a random choice of the two-body interaction, even with all of the above kinematic 
constraints the probability of getting a seniority preserving Hamiltonian is negligible
as well as the  conditions of
condensate stability (\ref{cond2}) are not necessarily satisfied.
Thus, although random systems exhibit trends
similar to  those
encountered in realistic nuclei  with pairing \cite{johnson00,talmi01},
the numerical studies show no enhancement of pairing
in the low-lying states of random Hamiltonians. 
It has been demonstrated \cite{mulhall}
that in the ground state wave function of a random Hamiltonian on a single $j$-level  
the $s=0$ pairing component
appears on a statistical level, i.e. with the same probability as 
any other component allowed by symmetries.  
Therefore it has been argued that the presence of regular pairing, a prominent
part of realistic physics, is not reproduced in randomly interacting systems
\cite{horoi01}. 
The
intrinsic feature of interactions describing  realistic systems is the
presence of  correlations between different interaction parameters.
These correlations, along with kinematic features, such as discussed above and 
other dynamic couplings, is what makes
the pairing effects survive and even dominate in the low-lying states of
many realistic nuclei.

\section{Pairing plus quadrupole model}
\label{SECPQ}
As discussed above, the most general
Hamiltonian can be separated into three parts, quasiscalar,
quasivector and a second rank tensor in quasispin.
The perturbation theory based on pairing treats exactly all
quasiscalar and quasivector components.
Since all odd multipoles are quasiscalars, only
those non-pairing interactions that can
be expressed in terms of the multipole operators of an even order,
starting from  $K=2\,,$ are of interest as the most orthogonal to pairing.
This leads to the  Hamiltonian
\begin{equation}
H=G \,P^{\dagger} P - \sum_{K=2,4\dots}\frac{\chi_K}{2} \sum_{\kappa} {\cal M}^{\dagger}_{K\,\kappa} 
{\cal M}_{K\,\kappa}\,.
\label{HPandM}
\end{equation}

The lowest possible $K=2$ multipole is responsible for quadrupole deformations and is
usually the most energetically favorable. Thus, we will further concentrate
on the pairing plus quadrupole (P+Q) 
Hamiltonian as defined below,
\begin{equation}
H=G \,P^{\dagger} P - \frac{\chi_2}{2} \sum_{\kappa} {\cal M}^{\dagger}_{2\,\kappa} 
{\cal M}_{2\,\kappa}\,.
\label{HPandQ}
\end{equation}
The physically relevant parameters correspond to attractive pairing
$G < 0 $ and
attraction in the quadrupole channel, $\chi_2>0\,.$ 
This Hamiltonian is
interesting for a number of reasons:
it accounts for both short- and long-range parts of the residual
interaction of nucleons through pairing and quadrupole parts, respectively, 
and consists of two very different components. Each of them  
separately is known to 
be responsible for 
collective phenomena, however acting simultaneously they lead to an 
interesting interplay. 
The study of  pairing versus deformation within P+Q model  
is usually carried out with Hartree-Bogoliubov (HB) technique \cite{baranger65}. In fact
the model is often defined as an arena for application of the HB method 
\cite{RingSchuck}, ignoring
exchange terms and previously discussed kinematic limitations. 
Studying the P+Q model beyond the HB approximation will be our further goal.

\subsection{Kinematic pairing}
\label{kinematic}
\begin{figure}
\begin{center}
\epsfxsize=8.0cm \epsfbox{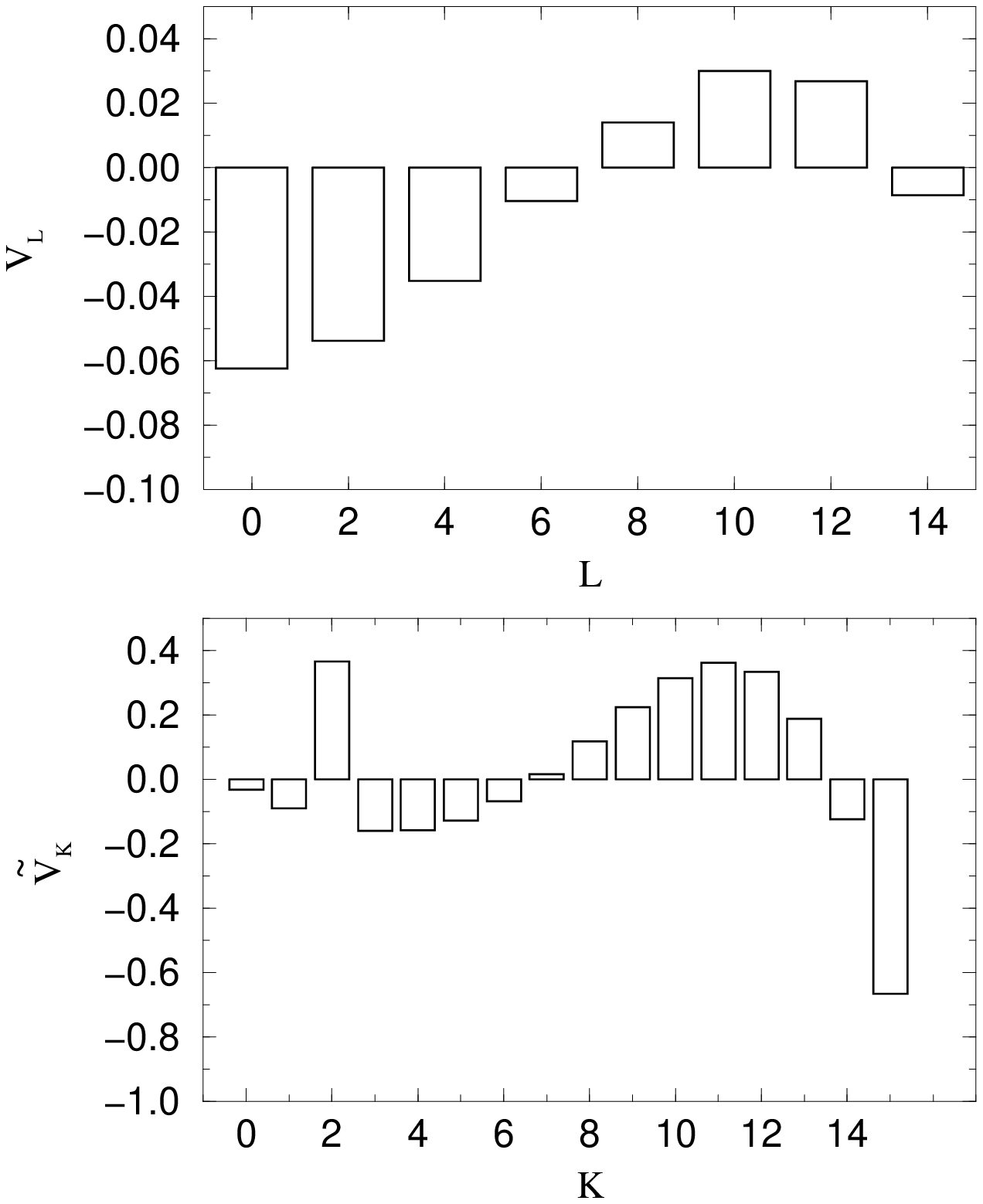}
\end{center}
\caption{Interaction parameters in the particle-particle channel (upper plot)
and particle-hole channel (lower plot) for a pure quadrupole-quadrupole
interaction, $\chi_2=1\,$ and $G=0$, in the model space $j=15/2$ are plotted as a
function of multipolarity. Notice different scales on the two panels. 
\label{qqV}}
\end{figure}

The interaction parameters of the Hamiltonian (\ref{HPandQ})
do not satisfy Eq. (\ref{onel_rel}) and,
as previously discussed,
this Hamiltonian
contains a non-physical part.
As a result the fact  that quadrupole part
contributes to pairing as well as to all other components in the $p-p$ or $p-h$ channel, 
and that pairing makes a contribution to the quadrupole part is not seen explicitly. 
In order to observe these kinematic couplings we will reduce the form of the Hamiltonian
(\ref{HPandQ}) with projection operators to a unique form where 
conditions (\ref{onel_rel}) are satisfied.
We rewrite the 
Hamiltonian (\ref{HPandQ}) in the following form 
\begin{equation}
H=\epsilon N+ \sum_L V_L \sum_{\Lambda} P^{\dagger}_{L\,\Lambda} P_{L\,\Lambda}
=\,-\sum_K \frac{\tilde{V}_K}{2} \sum_{\kappa} {\cal M}^{\dagger}_{K\,\kappa} 
{\cal M}_{K\,\kappa}\,,
\label{PandQC}
\end{equation}
removing the unphysical part with the projection operation,
\begin{equation}
\tilde{V}_K=\hat{\Theta}_K\,\chi_2 - (-)^K\,(2K+1) \frac{G}{\Omega}\,.
\end{equation}
Thus
\begin{equation}
\tilde{V}_K=\frac{\chi_2}{2}\,\delta_{K\,2}+ (2K+1)\,(-)^K\,\left (
\frac{\chi_2}{2}\, \left\{\begin{array}{ccc}
                               j&j&2\\
                               j&j&K\end{array}\right\} -\frac{G}{\Omega} 
\right )\,.
\end{equation}
In particular, the monopole part is 
\begin{equation}
\tilde{V}_0=-\frac{\chi_2+2G}{2\Omega}\,, 
\end{equation}
and there is a renormalization of the quadrupole strength which in the large
$\Omega$ limit behaves as
\begin{equation}
\tilde{V}_2=\frac{\chi_2}{2}-5 \,\frac{\chi_2+2G}{2\Omega}\,. 
\end{equation}
In the particle-particle channel we have
\begin{equation}
V_L=\delta_{L\,0}\,G+ \chi_2\,
\left\{\begin{array}{ccc}
                               j&j&2\\
                               j&j&L\end{array}\right\} \,,\quad \epsilon=
-\frac{\chi_2}{2 \Omega}\,.
\label{PandQV0}
\end{equation}
Eq. (\ref{PandQV0}) shows that $V_0=G-\chi_2/\Omega\,,$ which indicates that
even a pure attractive quadrupole interaction ($G=0$)
generates attractive pairing with the strength of the order $V_0\sim -1/\Omega\,.$ 
Furthermore, as follows from the properties of
$6j$-symbols, even in the case of $G=0$ pairing
is still the most attractive residual two-body force $V_0<V_{L\ne 0}\,.$
Therefore, in this model the pure attractive quadrupole-quadrupole Hamiltonian
has a paired ground state ($s=0$ and $J=0$) for nearly magic configurations
$N=2$ and $N=\Omega-2\,,$ this effect was also observed in other studies of
nucleon-phonon interactions \cite{kadmenskii87}.
A typical behavior of $V_L$ and $\tilde{V}_K$ as a function of $L$ and $K\,,$
respectively, is shown in Fig. \ref{qqV}.

The above expression indicates  that if $\Omega$ is very large then
only $V_0\approx G$ and $\tilde{V}_2 \approx \chi_2/2$ do not scale as 
$1/\Omega$, which leads to the usual P+Q model 
\cite{baranger65} with effectively decoupled quadrupole and pairing modes.
This is not surprising because  kinematic pairing as well as 
other kinematic couplings have mesoscopic geometry of nuclear systems as a 
source. 

In the P+Q model the condition for stability of pairing
in a nearly full or nearly empty shell, Eq. (\ref{cond1}),
is fulfilled even  without
any explicit pairing component $G=0\,,$ since as discussed above $V_0<V_{L\ne 0}\,.$  
However, an instability with the origin in the quadrupole
channel appears in the
middle of the shell,  where Eq. (\ref{cond2}) leads to the following
inequality in the limit of large $\Omega\,,$
\begin{equation}
G+\frac{\chi_2}{10}\,< 0\,.
\label{cond3}
\end{equation}
For the general case
of Hamiltonian (\ref{HPandM}),
the condition (\ref{cond3})
becomes
\begin{equation}
G+\frac{\chi_K}{2(2K+1)}\,<
0\,,\quad K\,\,{\rm is \,\,even}\,.
\label{condX}
\end{equation}
From the properties of the $6j$-symbols it follows that the 
kinematic pairing resulting
from any attractive multipole-multipole (even multipolarity) interaction is
always attractive and is the strongest two-body 
component $V_0<V_{L\ne 0}\,$ in the $p-p$ channel.
The above result also indicates that lower 
multipoles ($K=2$ is the lowest one)
are more likely to destroy pairing because of the suppression factor
$1/(2K+1)\,.$

For completeness we present an exact equation for the separation energy of a
pair from the condensate which is the minus
excitation energy of the
first  $J^{+}$ state with seniority $s=2$ for the model defined by the
Hamiltonian 
(\ref{HPandM}),
\begin{equation}
S=G+B_J\,.
\end{equation}
Here $B_J$ is independent of $G$ ($J\ne 0$) and  given by equation
$$
B_J=-V_J-(N-2)\frac{Y}{2}+\frac{(N-2)(\Omega-N-2)}{(\Omega-6)(\Omega-4)}
\left \{ (\Omega-8) Y + 4 V_J + \frac{2 \chi_J}{2J+1} \right \}\,
$$
$$
-\frac{N Y}{2(\Omega-2)} \,(\Omega-2N+2)\,,
$$
where
\begin{equation}
Y=\frac{1}{\Omega}\,\sum_{K=2,4,6\dots} \chi_K=G-V_0\,,
\end{equation}
and $V_J$ are determined as
$$
V_J=\delta_{J\,0}\,G+ \sum_K\,\chi_K\,
\left\{\begin{array}{ccc}
                               j&j&K\\
                               j&j&J\end{array}\right\} \,.
$$

\begin{figure}
\begin{center}
\epsfxsize=8.0cm \epsfbox{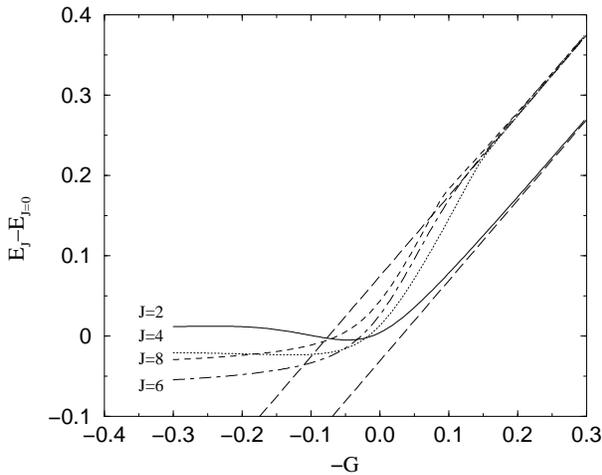}
\end{center}
\caption{
The spectrum of lowest even-spin states relative to the lowest spin zero state  
in the system of six particles on $j=15/2$ orbital for the P+Q interaction as a function of $G\,.$ Two dashed straight 
lines that correspond to
perturbation theory $E_2-E_0=-G-B_2\,$ and $E_4-E_0=-G-B_4\,$ are plotted
for comparison (in this model $B_2=0.031$ and $B_4=-0.074$).
The quadrupole strength is set
at $\chi_2=1\,.$
\label{PQSPECTR}}
\end{figure}

\begin{figure}
\begin{center}
\epsfxsize=8.0cm \epsfbox{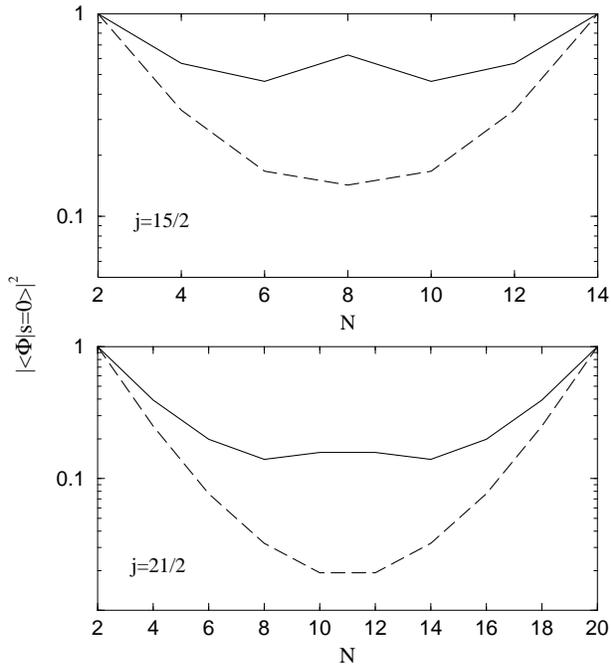}
\end{center}
\caption{One-level systems with $j=15/2$ and $j=21/2\,,$ upper and lower plots,
respectively, are considered with pure quadrupole-quadrupole interaction,
$G=0\,.$ The wave function of the spin zero ground state is overlapped with
the spin zero ground state of a
paired system ($G=-1$ and $\chi_2=0$), i.e. with the
seniority $s=0$ state, that is unique for a single $j$-level system.
The square of this overlap is plotted in solid line as a function of a
particle number. This result is compared to the  
statistical expectation of pairing from random interactions, shown by dashed
curve. The statistical expectation is defined here as an inverse number of
spin zero states in the system [2].
\label{PQoverlap}}
\end{figure}

With pairing as the only interaction, the lowest excited state in the system with
seniority $s=2$ is at  two-quasiparticle excitation
energy, which is $G$ in this case.
Other interactions can lower this energy
by a constant $B_J$ which is mainly effected by the term $\chi_J/(2J+1)\,.$
Since we are dealing with the zeroth order
perturbation theory, this gap between the ground
state and a lowest excited state of a given spin behaves linearly with the pairing strength.
For example, in the P+Q model (see Fig. \ref{PQSPECTR})
the state $J=2$ is the
lowest excited state in the paired region. 
The pairing based description for the system of six particles on the $j=15/2$
level becomes unstable for
$G\ge-0.1\,$ (negative $G$ corresponds to attraction in the 
pairing
channel) as predicted by Eq. (\ref{cond3}) with
$\chi_2=1\,;$ this agrees well with the comparison presented in 
Fig. \ref{PQSPECTR}\,. However, in the region where unperturbed paired states
can not be used to describe the system, such as the extreme
case of  no explicit pairing, $G=0\,,$ the
effects of kinematic pairing are still quite strong. 
In order to emphasize this,
the numerical values of overlaps between the ground states of P+Q 
systems with no explicit pairing ($G=0\,,\,\, \chi_2=1$) and paired
systems ($G=-1\,,\,\, \chi_2=0$) are shown in Fig. \ref{PQoverlap}. These results
indicate significant pairing, that greatly exceeds the statistical prediction
relevant to random interactions \cite{mulhall}, shown in the figure by dashed 
lines.
The enhancement of pairing in the middle of the shell, observed in both plots
of Fig. \ref{PQoverlap}, is related to another kinematic feature. As it
was discussed, 
exactly in the middle of the shell, due to the particle-hole symmetry the 
deformation must disappear which allows for more pairing correlations. In the following
subsection we will further discuss this issue.

\begin{figure}
\begin{center}
\epsfxsize=8.0cm \epsfbox{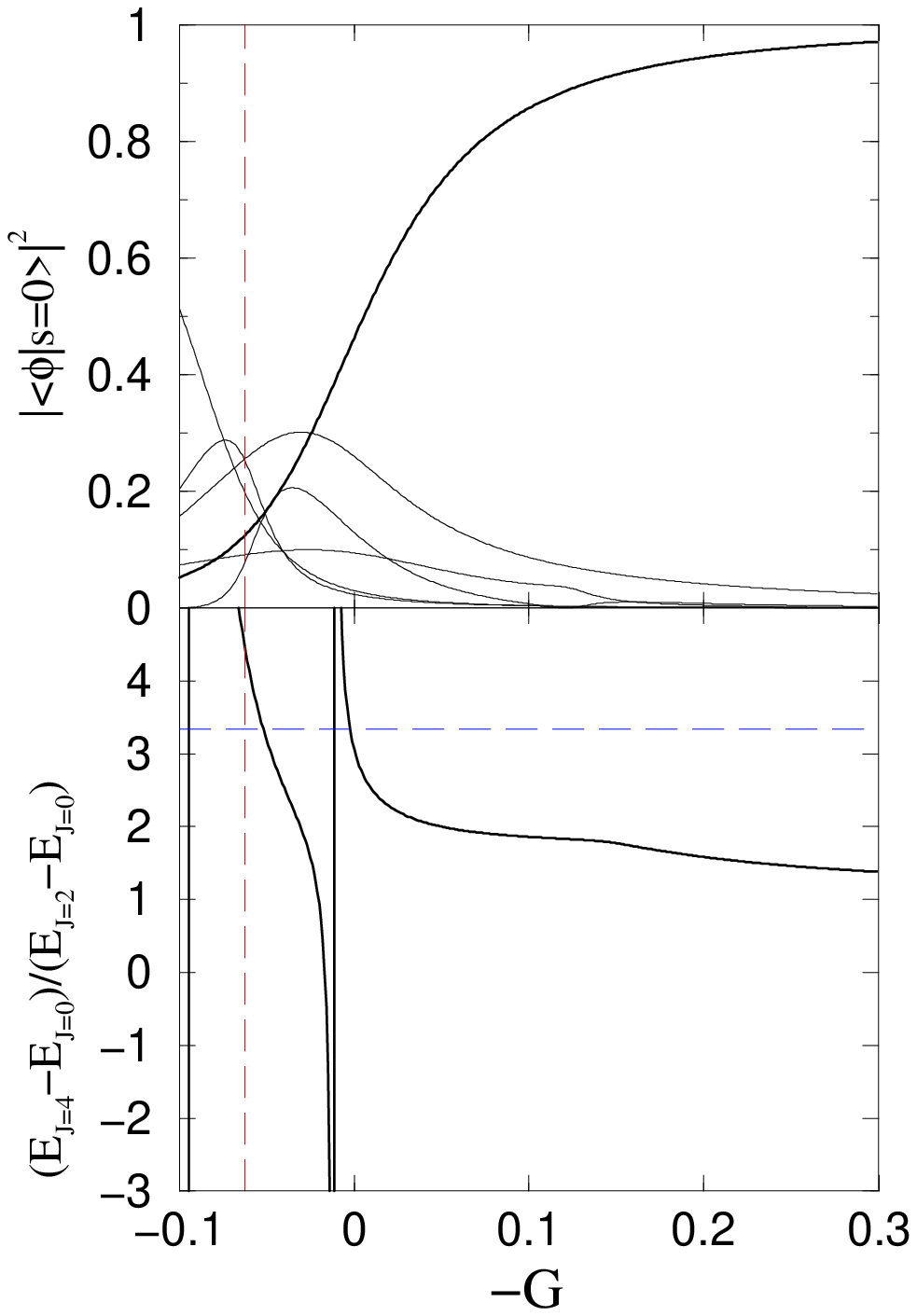}
\end{center}
\caption{Properties of the system of six particles on $j=15/2$ orbital with the
P+Q interaction are
studied as a function of the parameter $G\,;$ the quadrupole strength is set
at $\chi_2=1\,.$ The upper plot shows the overlap of all six $J=0$ eigenstates in
this system with the $s=0$ pairing state, which is defined as a ground state
wave function for the pairing interaction $G=-1\,$ and $\chi_2=0\,.$
Lower plot shows the ratio $(E_{J=4}-E_{J=0})/(E_{J=2}-E_{J=0})$ as a function
of $G$, here $E_{J=4,2,0}$ are the energies of the lowest states with spin 4, 2, and 0,
respectively. 
\label{PandQ6}}
\end{figure}

The same result is seen from the upper plot of Fig. \ref{PandQ6} which shows that 
the pure quadrupole-quadrupole interaction, 
the region of $G=0\,,$ has a ground state
dominated by the $s=0$ pairing component. 
Only in the region where kinematic pairing is explicitly
balanced (this point $V_0=0$ is shown by the dashed vertical line) by 
the repulsive explicit pairing given via positive $G\,,$ the
preference to $s=0$ pairing component
disappears, and all $J=0$ states have almost the same overlap with $s=0$ 
wave function. The lower plot of Fig. \ref{PandQ6} addresses the low-lying
properties of the spectrum in the same region of parameters; it will
be discussed in the following subsection.

\subsection{Pairing and deformation}
\label{pandd}
In this subsection we will investigate the interplay of
pairing
and deformation in the P+Q model.
Rotational bands in
the low end of the spectrum, 
and the Alaga intensity rules will serve us
as tools in this study. 
The lower part of Fig. \ref{PandQ6} shows
that an indication for a rotational band near the ``no pairing''
region, $G=0\,.$ Here, judging by the excitation energy, the lowest states with $J=0,2$ and 4 
are forming a collective rotational band. 
As it is clear from the figure this region is 
very small. Thus, for the most part we
will concentrate here on a pure quadrupole-quadrupole interaction
\begin{equation}
H=-\frac{\chi_2}{2} \sum_{\kappa} {\cal M}^{\dagger}_{2\,\kappa} 
{\cal M}_{2\,\kappa}\,,
\label{QQHAM}
\end{equation}
which, as
we have shown above, is still
significantly influenced by the kinematic pairing, with
the bandhead $J=0$ state 
dominated by the $s=0$ pairing component. 

We note that although single-$j$ model is very limited, an exact rotational
spectrum can still be formed using
\begin{equation}
H=\chi_1 \sum_{\kappa} {\cal M}^{\dagger}_{1\,\kappa} 
{\cal M}_{1\,\kappa}=\frac{\chi_1}{j(j+1) \Omega} J^2 \,,
\end{equation}
which is seniority conserving and,  according to Eq. (\ref{onel_vl})
can be defined in the $p-p$ channel with the aid of the set of parameters 
\begin{equation}
V_L=L(L+1)-2j(j+1)\propto \left \{\begin{array}{ccc}
j&j&L\\
j&j&1\end{array}\right\}\,.
\label{mom}
\end{equation}
This interaction remarkably well satisfies the Alaga intensity rules, see
below,  for
reasonably large $j$ and $N\,.$
In the model with random interactions \cite{mulhall,yadfiz},
the contributions 
(\ref{mom}) determine the effective statistical moment of inertia.
Similar to this example other odd-multipolarity multipoles can be used
to create seniority conserving interactions, this is another way of
finding solutions to Eq. (\ref{nomix}).

The main hint for the presence of deformation comes from the mean
field approximation
\cite{baranger65} (with exchange terms ignored by definition of the model this is in fact
a Hartree approximation).
For  axially symmetric deformation, the average values of the multipole
moments
are  
\begin{equation}
\langle {\cal M}_{2\,0}\rangle=\sum_{m} \frac{2 (3m^2-j(j+1))}
{\sqrt{\Omega(\Omega^2-1)(\Omega^2-2)}} \, n_m
\,,
\end{equation}
in terms of the occupation numbers $ n_m=\langle a^\dagger_m\,a_m\,\rangle $ 
in the intrinsic frame with the $z$-axis oriented along the symmetry axis,
and
$$
\langle {\cal M}_{2\,-2}\rangle=\langle {\cal M}_{2\,2}\rangle=0\,.
$$
The single-particle energies in the body-fixed frame  can be obtained
via the usual self-consistency requirement
\begin{equation}
\epsilon_m= -\chi_2\,\,\frac{2 (3m^2-j(j+1))}
{\sqrt{\Omega(\Omega^2-1)(\Omega^2-2)}}\,\langle {\cal M}_{2\,0}\rangle\,.
\label{HFSP}
\end{equation}
This problem
can be solved exactly \cite{baranger65}.
The minimum energy corresponds to the 
Fermi occupation of the $N$ lowest pairwise degenerate orbitals
$|m|=1/2,3/2,\dots (N-1)/2$ for prolate or
$|m|=j,j-1,\dots j-(N-2)/2$  for oblate shapes.
The corresponding quadrupole moment is  then  given as
\begin{equation}
\langle {\cal M}_{2\,0}\rangle =-\frac{1}{4}\,\frac{N(\Omega^2-N^2)}
{\sqrt{\Omega(\Omega^2-1)(\Omega^2-4)}}
\end{equation}
for prolate deformation and
\begin{equation}
\langle {\cal M}_{2\,0}\rangle =\frac{1}{4}\,\frac{N(2\Omega-N)(\Omega-N)}
{\sqrt{\Omega(\Omega^2-1)(\Omega^2-4)}}
\end{equation}
for oblate deformation.
The deformation energy, defined as 
\begin{equation}
E_{\rm deformation}=-\frac{\chi_2}{2}\,|\langle {\cal M}_{2\,0}\rangle|^2\,,
\end{equation}
shows that 
oblate deformation is preferred for $N<\Omega/2$ and prolate
correspondingly for $N>\Omega/2\,.$
In the middle of the shell there is a phase transition
from prolate to oblate deformation. In this region energies associated 
with prolate and oblate deformation 
become equal. The average quadrupole moments at this point do not vanish; they 
are opposite in sign for prolate and oblate deformation.
Because of this phase transition, mesoscopic nature of the system,
and all other kinematic terms ignored in the model,
the true ground state is a superposition of 
oblate and prolate forms in the region of a half-occupied shell. 
At the exact point of half-occupancy the particle-hole symmetry requires 
$\langle {\cal M}_{2\,0}\rangle=0\,$ for the true ground state of the system. 
The oblate to prolate 
transition turns out 
to be advantageous to pairing, as can be seen from Fig. \ref{PQoverlap}. The
content of pairing, the $s=0$ component in the wave function, is slightly
enhanced in the
middle of the shell. This enhancement, being accompanied by strong effects
of kinematic 
pairing near both limits $N=0$ and $N=\Omega\,,$ is largely responsible for the
presence of pairing correlations  throughout the entire region
within a single shell.

The moment of inertia in the cranking approximation,
that will be of use in our further 
discussion, is given by the following expression
\begin{equation}
{\cal I}=-\sum_{m\,m'} \frac{n_m-n_{m'}}{\epsilon_m-\epsilon_{m'}}\,|j_x|^2\,.
\end{equation}
With a sharp Fermi surface only four terms in the sum, $|m|=|m'|\pm 1$ will work.
Utilizing Eqs. (\ref{HFSP}) and (\ref{M1}) we obtain
\begin{equation}
{\cal I}_{\rm
prolate}=\frac{\Omega(\Omega^2-1)(\Omega^2-4)}{6\,\chi_2\,N^2}\,,\quad
{\cal I}_{\rm oblate}=\,\frac{\Omega(\Omega^2-1)(\Omega^2-4)}{6\,\chi_2\,(\Omega-N)^2}\,.
\label{HFi}
\end{equation}

The Elliot's SU(3) model 
has  a clear rotational structure and  can serve as a
link for understanding a macroscopic deformation and its microscopic description.
In the 
quadrupole-quadrupole Hamiltonian under consideration  the SU(3) 
algebra is broken 
in such a kinematic way
that boosts competing  pairing.
The commutator of the quadrupole operators, according to Eq. (\ref{algmult}),
consists of vector and octupole components
\begin{equation}
[{\cal M}_{2\,\kappa},\, {\cal M}_{2\,\kappa'}]=
6\,\sum_q\,(-)^q\, \left \{\begin{array}{ccc}
2&2&1\\
j&j&j\end{array}\right\} \,\left (\begin{array}{ccc}
2&2&1\\
\kappa&\kappa'&-q\end{array}\right)\,{\cal M}_{1\,q}
\end{equation}
$$
+
14 \,\sum_q\,(-)^q\, \left \{\begin{array}{ccc}
2&2&3\\
j&j&j\end{array}\right\} \,\left (\begin{array}{ccc}
2&2&3\\
\kappa&\kappa'&-q\end{array}\right)\,{\cal M}_{3\,q}\,.
$$
The octupole component is what breaks the SU(3) algebra; ignoring this term,
and similar to Eq. (\ref{M1}), introducing a quadrupole operator according to
\begin{equation}
{\cal M}_{2\,q}=\frac{{ Q}_q}{\sqrt{5\,j(j+1)\,\Omega}}\,,
\end{equation}
we obtain a SU(3) algebra in the standard form
\begin{equation}
[J_q\,,J_{q'}]=-\sqrt{2}\,C_{1 q,\,1 q'}^{1 q+q'}\,J_{q+q'}\,
\end{equation}
\begin{equation}
[{ Q}_q\,,J_{q'}]=-\sqrt{6}\,C_{2 q,\,1 q'}^{2 q+q'}\,{ Q}_{q+q'}\,
\end{equation}
\begin{equation}
[{ Q}_q\,,{ Q}_{q'}]=3\sqrt{10}\,C_{2 q,\,2 q'}^{1 q+q'}\,J_{q+q'}\,.
\end{equation}
The quadrupole-quadrupole Hamiltonian can be expressed via the bilinear Casimir
operator of this group
\begin{equation}
{\cal C}=Q\cdot Q+ 3\, J \cdot J\,.
\end{equation}
The expectation value of this invariant operator depends only on quantum numbers
$\lambda$ and $\mu$ that label representations, see for example \cite{harvey},
\begin{equation}
\langle {\cal C} \rangle = 4(\lambda^2+\mu^2+\lambda\,\mu+ 3\lambda + 3\mu )\,.
\end{equation}
For a given representation $(\lambda\,\mu)\,,$ where $\lambda\ge \mu\,,$ the angular
momentum can be
\begin{equation}
L=\left \{ \begin{array}{lc} K,K+1,K+2,\dots K+\lambda & K\ne 0\,,\cr
\lambda,\lambda-2\dots (1\,\,{\rm or}\,\,0) & K=0\,,\cr
\end{array}\right .\,
\end{equation}
where an integer $K$ takes values $K=\mu,\mu-2,\dots (1\,\,{\rm or}\,\,0)\,.$
Assuming a deformed band based on the ground state  $K=0$ 
we choose $\mu=0$ and $\lambda=N(\Omega-N)/2\,$ 
where $\lambda$ coincides with the maximum possible  value of angular momentum
in the system.
Thus in the approximation of SU(3) symmetry the expectation value of the
quadrupole-quadrupole Hamiltonian becomes
\begin{equation}
\langle N\,(J) | H |N\,(J)\rangle =-\frac{2\,\chi_2\,N(\Omega-N)(N\Omega-N^2+6)}
{5\,\Omega(\Omega^2-1)}+\frac{6\,\chi_2\, J(J+1)}{5\,\Omega(\Omega^2-1)}
\end{equation}
The SU(3) model results in exact rotational bands with the $N$-independent
moment of inertia 
\begin{equation}
{\cal I}=\frac{5\Omega (\Omega^2-1)}{12 \chi_2}\,.
\label{su3i}
\end{equation}

\begin{figure}
\begin{center}
\epsfxsize=8.0cm \epsfbox{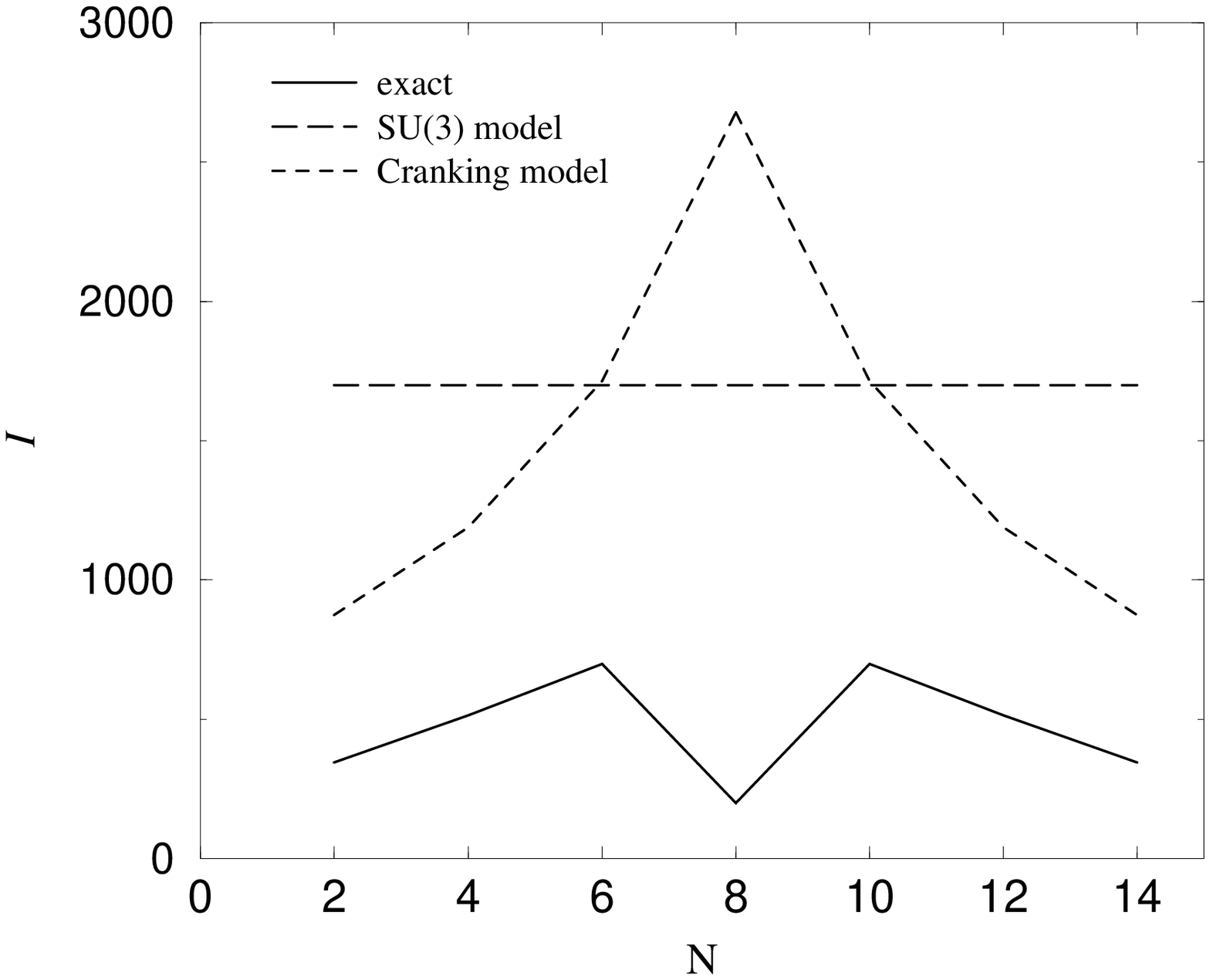}
\end{center}
\caption{Moments of inertia obtained using different models and
from fitting  the low-lying states in the exact spectrum 
are compared as a function of a number of particles.
The pure quadrupole-quadrupole
interaction is used ($\chi_2=1$ and $G=0$) for the system on a
single $j=15/2$ level.
The solid line corresponds to the moment of inertia coming from 
fitting the exact spectrum, short-dashed line corresponds to the moment
of inertia from the mean field treatment, and long dashed line is for the 
SU(3) model, the latter predicts an $N$-independent value.
\label{PQmomentj15}}
\end{figure}

In Fig. \ref{PQmomentj15} the moments of inertia obtained from Eqs. 
(\ref{HFi}) and (\ref{su3i}) and from fitting the $J=0,2,4$ states
in the spectra obtained from exact diagonalization are compared.
In a view of the previous discussion 
it is not surprising that observed moments of inertia are
significantly lower than ones predicted by both of the considered models. 
This  effect 
should be attributed to pairing, which is mainly of a  kinematic 
origin. 

Static deformation of a nucleus manifests itself via relations 
(the so-called Alaga intensity rules)
between the 
diagonal expectation values of multipole moments and off-diagonal, 
transitional amplitudes corresponding to the same $K$ band.
We consider here a quadrupole moment of the lowest excited $J=2$ state
and the E2 transition between this state and the $J=0$ ground state.
In a single level model the quadrupole operator must be proportional to ${\cal
M}_{2\,\kappa}$ as it is the only particle-hole operator with the correct
rotational properties. 
We define 
quadrupole moment of the state  as
\begin{equation}
{\cal Q}=\langle J M=J|{\cal M}_{2\,0}|J\, M=J\rangle\,.
\end{equation}
The B(E2) transition strength is defined as
\begin{equation}
{\rm B(E2)}=\sum_{M_f\,\kappa}\,|\langle J M_f|{\cal M}_{2\,\kappa}|J\, M_i|^2\,.
\end{equation}
The Alaga intensity rules are then expressed via relation \cite{zelevinsky,BM}
\begin{equation}
\frac{{\cal Q}^2}{\rm B(E2)}=\frac{4}{49}\,.
\label{ala}
\end{equation}

Within the pairing based treatment of interactions which implies no seniority mixing, 
this ratio can be calculated 
analytically. 
We assume that the lowest $J=2$ state has pure seniority $s=2\,.$ 
Due to the seniority conservation, this is the only state that can directly
decay into the ground state, assumed here to have $s=0\,,$ via an E2 transition.
The amplitude of this transition is
given by
\begin{equation}
\langle s=2,K \kappa|{\cal M}^{\dagger}_{K\,
\kappa}|s=0\rangle=-\sqrt{\frac{2N(\Omega-N)}{(2K+1)\Omega (\Omega-2)}}\,,
\end{equation}
therefore in the limit of strong pairing 
\begin{equation}
{\rm B(E2)}=\frac{2N(\Omega-N)}{\Omega (\Omega-2)}\,.
\label{EPBE2}
\end{equation}
The rate of this transition is maximized for a half-occupied system.
The corresponding quadrupole moment in the $J=2$ and  $s=2$ state
can be determined from
\begin{equation}
\langle N, s=2, (J M)|{\cal M}_{K\,0}|N, s=2, (J M) \rangle
\end{equation}
$$= 2(2J+1)\,\frac{2N-\Omega}{\Omega-4}\,(-)^M
\left (\begin{array}{ccc}
                               J&K&J\\
                               M&0&-M\end{array}\right)
\left\{\begin{array}{ccc}
                               J&K&J\\
                               j&j&j\end{array}\right \}\,.
$$
Application for quadrupole $K=2$ results in the following expression
\begin{equation}
{\cal Q}=\frac{4}{7}\,\frac{(2N-\Omega)(\Omega+4)}{\sqrt{\Omega(\Omega^2-1) (\Omega^2-4)}}\,.
\label{EPQ}
\end{equation}
The quadrupole moment goes to zero for the half-occupied shell as required by
the particle-hole symmetry. 

\begin{figure}
\begin{center}
\epsfxsize=8.0cm \epsfbox{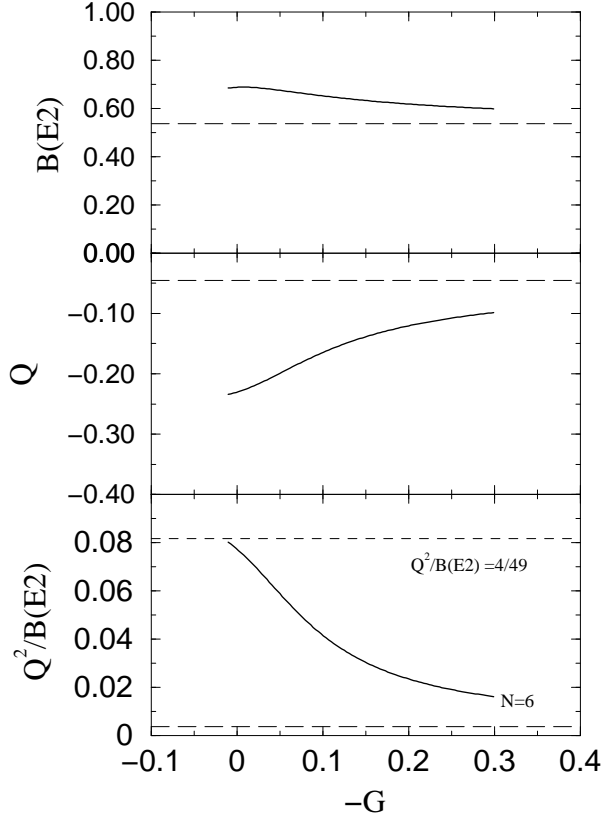}
\end{center}
\caption{The quadrupole characteristics of the first excited $2^{+}$ state, 
the diagonal quadrupole moment $Q$, Eq. (97), middle panel, and the reduced transition
probability to the ground state, B(E2), Eq. (98), upper panel, in the 
P+Q model for the single $j=15/2$
level and $N=6\,,$ as a function of pairing strength $G\,.$
Long dashed line on all plots indicates the results from pairing based
description.
The lower plot shows the prediction of the 
Alaga intensity rule $Q^2/{\rm B(E2)}=4/49\,,$ short dashed line.
\label{alag}}
\end{figure}

\begin{figure}
\begin{center}
\epsfxsize=8.0cm \epsfbox{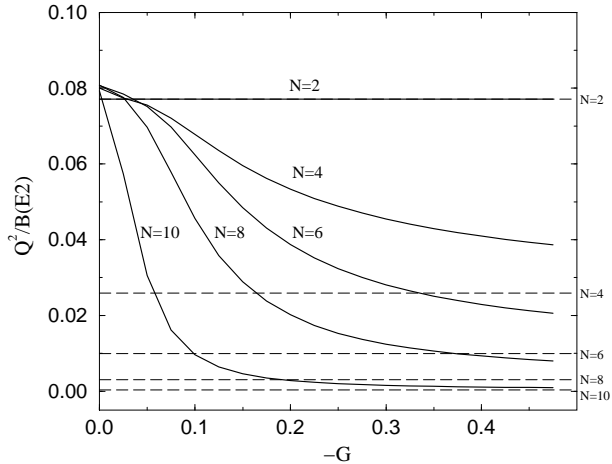}
\end{center}
\caption{Ratio $Q^2/{\rm B(E2)}$ for systems with different numbers of particles $N$ on a single
$j=21/2$ level is plotted in solid lines as a function of pairing strength $G\,.$ 
Thin dashed lines, with corresponding values of $N$ marked on the right,
indicate the intensity ratios for
corresponding systems in the limit of validity of pairing perturbation theory.
All results should be compared with $Q^2/B(E2)=4/49\approx0.082\,,$ the
Alaga intensity ratio of a rotating rigid body.
\label{alaga2}}
\end{figure}

In general, the paired state is not deformed and thus the B(E2) transition probability in 
(\ref{EPBE2}) and the quadrupole moment (\ref{EPQ}) do not satisfy the intensity
rule (\ref{ala}). With the quadrupole-quadrupole interaction presented, a deformation can 
appear and Alaga intensity rules  can be fulfilled.
In Fig. \ref{alag} for the system $j=15/2$ and $N=6\,,$ the quadrupole moment,
B(E2) transition strength, and the ratio ${\cal Q}^2/{\rm B(E2)}$ are presented as 
a function of the pairing strength $G\,.$ Dashed lines show the results 
for the strong pairing limit, obtained using  Eqs.  (\ref{EPBE2}) and
(\ref{EPQ}).  
Fig. \ref{alaga2} compares the behavior of the ratio $Q^2/{\rm B(E2)}$ as a function
of the pairing strength $G$ in different systems $N=2,4,6,8,10$ with  
$j=21/2\,.$ 
There are two special cases.  For $N=2$ 
(the same is true for $N=\Omega-2\,$)  
the ground state is paired, and the Alaga ratio is determined exactly 
via Eqs.  (\ref{EPBE2}) and (\ref{EPQ}). The second case is for $N=\Omega/2\,,$ 
where due to particle-hole symmetry  $Q^2/{\rm B(E2)}=0\,.$ For all other situations
Alaga intensity rules are fulfilled to a sufficient degree of accuracy 
for the pure quadrupole-quadrupole 
Hamiltonian $G=0\,.$ 
As the pairing strength increases, the ratio  $Q^2/{\rm B(E2)}$ moves
away from the Alaga value towards the limit determined by pairing,
which is shown 
by thin dashed lines.  Curves corresponding to systems with a number of 
particles close to $N=2\,$  ($N=\Omega-2\,$) or $N=\Omega/2\,$ converge to 
the pairing limit significantly faster. This fact is yet another manifestation
of preference for pairing correlations in these systems.  

\section{Conclusions}
The pairing interaction is a very important part of the general residual
interaction. However the fact that many nuclei are paired in their ground or
low-lying exited states is a result of a non-trivial interplay of pairing matrix
elements and other parts of the residual interaction, 
as well as kinematic constraints. 
It has been earlier shown through a number of numerical studies that the
realistic interaction
is different from an arbitrary symmetry preserving random interaction.
The difference lies in the correlations between the interaction parameters that
exist even in the truncated shell model space, reflecting the
overall properties of the nuclear medium. Interactions that permit paired
states and allow for deformations produce very correlated sets of
residual two-body matrix elements. Apart from the dynamic correlations
there are  kinematic
couplings between different collective modes, that mainly appear from the geometrical
restrictions imposed on collective excitations. 

In many  realistic
systems pairing is relatively weak compared to the critical value of
phase transition defined by BCS. However, as
was shown by numerous authors, pairing
correlations in mesoscopic systems extend far beyond the BCS phase transition.
This makes the exact treatment of pairing a necessary component in the study
of a sensitive interplay between pairing and other interactions.

The simple single-level systems considered in this work has served  as 
an excellent testing ground. 
The SU(2) quasispin algebra allows for an exact solution of the pairing
problem, and perturbation theory based on the 
paired state \cite{EP} can be further developed with ease.
This makes it possible to address important 
questions of the stability of the paired state as well as to investigate 
dynamic renormalizations of two-body interactions in the presence
of the pairing condensate.
 
The interaction parameters for a one-level system
in the particle-particle and particle-hole channels can be  
related to each other, revealing kinematic constraints in a simple form.
Additional restrictions on the behavior of the system come from the 
exact particle-hole symmetry.  As a result of this symmetry, all
multipole moments of even multipolarity are identically zero for 
a half-occupied shell. 
These 
constraints turned out to be very important for the preservation of
pairing in the presence of deformation. 
Interplay of pairing and deformation was discussed using the P+Q model.
The particular case of this model, 
the pure quadrupole-quadrupole interaction with no explicit
pairing, has been addressed in detail. 
This interaction was expected to be the one most ``orthogonal'' to pairing;
however, strong evidence of pairing correlations was found even in this case. 
We have shown that the pairing is the strongest attractive component in the
quadrupole-quadrupole interaction when the latter 
is converted from the $p-h$ to $p-p$ channel.
This means that for any attractive, i.e. favoring deformation, 
quadrupole-quadrupole interaction on a single 
$j$-level, the ground state of a two particle or two-hole system 
is a paired state ($J=0$ and $s=0\,$).
The same is in fact true for a more general Hamiltonian containing any 
attractive multipole-multipole
interaction of even multipolarity. 
This results in enhanced pairing 
correlations in near-magic configurations. 
An additional enhancement
of pairing was observed in the systems close to the half-filled shell. 
This is related to the prolate-oblate phase transition taking place in 
this region; it weakens 
deformation thus allowing for more pairing.
A number of calculations was performed to emphasize the
discussed aspects, and the results were compared to the predictions 
of the mean field approximation
and Elliot's SU(3) model.
Direct overlaps with the $s=0$ paired wave-function, excitation spectrum, 
intensities of transitions and moments of inertia all indicate an appearance 
of pairing, consistent with the kinematic constraints.    

The simple model used in this work is only a very limited 
approximation to the situation in nuclear systems. However, the
observed effects of pairing enhancement via kinematic or dynamic interplay
with other interactions have to exist to some extent in realistic 
systems. The situation in the  realistic shell model   
becomes much more diverse as it is complicated by 
an increasingly large number 
of interaction parameters, weakening of antisymmetry requirements and other kinematic 
restrictions, 
and a diversity of  collective modes.
Pairing
itself becomes different, isovector and isoscalar pairing modes can compete,
the 
pairing state of the lowest total seniority $s=0$ is not unique and can
involve coherent and incoherent 
pair vibrations \cite{volya01} which, under effects of other interactions, may
result in a paired condensate that is different from the prediction of  
the regular BCS.\\
\\
{\bf Acknowledgements:} The author wishes to thank V. Zelevinsky, B. A. Brown
and D. Mulhall for motivating discussions and useful criticism, 
without their help this work
would not be possible. The NSF support of this research is greatly
appreciated.

\end{document}